\documentclass[12pt,preprint]{aastex}
\begin{document}

\title{A Low Upper Limit to the Lyman Continuum Emission of two 
galaxies at \lowercase {$z\simeq 3$}\altaffilmark{1}}

\altaffiltext{1}{Based on observations
obtained in service mode at the ESO VLT for the program 65.O-0631}

\author{E. Giallongo}
\affil{Osservatorio Astronomico di Roma, via Frascati 33,
    Monteporzio, I-00040, Italy}
\author{S. Cristiani}
\affil{ECF, European Southern Observatory
    Karl-Schwarzschild Strasse, Garching bei Munchen, D-85748, Germany}
\affil{Osservatorio Astronomico di Trieste
    via G. B. Tiepolo, 11, Trieste I-34131 Italy}
\author{S. D'Odorico}
\affil{European Southern Observatory, Karl-Schwarzschild Strasse,
    Garching bei Munchen, D-85748, Germany}
\author{A. Fontana}
\affil{Osservatorio Astronomico di Roma, via Frascati 33,
    Monteporzio, I-00040, Italy}
\begin{abstract}

Long exposure, long-slit spectra have been obtained in the UV/optical bands  for
two galaxies at  $z=2.96$ and  $z=3.32$ to investigate the fraction  of ionizing
UV photons escaping  from high redshifts  galaxies.  The two  targets are  among
the brightest  galaxies  discovered   by Steidel  and   collaborators   and they
have  different  properties  in   terms   of Lyman-$\alpha$  emission  and  dust
reddening. No significant  Lyman  continuum  emission has   been  detected.  The
noise   level  in    the   spectra       implies     an     upper      limit  of
$f_{rel,esc}\equiv     3   f(900)/f(1500)<    16$\%   for   the  relative escape
fraction of ionizing photons, after correction for absorption by the intervening
intergalactic  medium.    This   upper      limit       is       4         times
lower       than the   previous  detection  derived     from      a    composite
spectrum       of 29  Lyman  break  galaxies  at   $z\simeq   3.4$.    If  this
value  is    typical of  the escape   fraction      of    the     $z\sim      3$
galaxies,    and    is  added  to     the  expected     contribution   of    the
QSO     population,    the  derived          UV  background    is   in      good
agreement   with   the    one derived   by    the proximity effect

\end{abstract}

\keywords{galaxies: distance and redshifts --- galaxies: formation --- 
intergalactic medium}

\section{Introduction}

The  integrated  ultraviolet  background (UVB)  arising  from  quasars and  hot,
massive stars   in star-forming  galaxies is likely responsible  for maintaining
the high  degree   of  ionization  of   the  intergalactic   medium  (IGM)    we
observe   in the    Lyman-$\alpha$  forest  lines  in the    absorption spectra
of background  QSOs  (e.g. Bechtold  1994; Giallongo et  al. 1996).  Moreover it
can  affect the  processes of galaxy  formation and evolution  in particular for
the less massive systems  (e.g. Babul \& Rees 1992; Benson et al. 2001).

The contribution of QSOs to the UVB as a function of redshift is the  easiest to
assess with some  degree of reliability  based on the  estimates of the   quasar
luminosity function from $z=0$ to $z\sim 5$ (e.g. Haardt \& Madau 1996).

Hot,  massive  stars  in  star-forming  galaxies  have  also  been  suggested as
important contributors  to the  UVB (Bechtold  et al.  1987; Songaila, Cowie, \&
Lilly 1990; Miralda-Escud\'e  \& Ostriker 1990; Bianchi, Cristiani, \& Kim 2001)
at  early  epochs.     Photometric  and  spectroscopic  data   on  the    galaxy
population at  $z\geq  3$   are increasing  rapidly   (Steidel    et  al.  1996;
Lanzetta    et   al.   1996; Giallongo   et   al.  1998;  Fontana      et    al.
1999,   2000)    and      are  providing  statistical  information    on     the
galaxy      luminosity     and  spatial distributions (Steidel et     al.  1999,
Poli     et     al.     2001).   Intermediate  resolution    spectroscopy     of
the   brightest      fraction   of  the     high  redshift    galaxies  has been
effectively used to derive  unique    information on the     physical properties
of      these     stellar   systems     like   the star  formation rate,     gas
dynamics,   metallicity,  and     dust content (e.g.  Pettini et al. 1998).

The detection by Tytler et al. (1995) and  by Cowie et al. (1995) of CIV in  the
Lyman-$\alpha$  forest  clouds has  provided  the first  evidence  of widespread
chemical enrichment in the IGM at $z\sim 3$. Madau \& Shull (1996) have computed
the ionizing stellar radiation flux  which accompanies the production of  metals
at high-$z$, thus  realizing  that  this   may be  significant,  comparable   to
the QSO contribution if a fraction  $\gtrsim 25$\% of the  UV radiation  emitted
from stars can escape into the intergalactic space. 

As a consequence, the importance  of the contribution  to the UVB  by the galaxy
population depends on  the ionizing  escape fraction  of the  UV photons   which
is a poorly known  parameter.   At low  redshifts,  Giallongo, Fontana, \& Madau
1997  and Shull et  al.  (1999) have  emphasized that  the  galaxy  contribution
is  already comparable  to the QSO contribution for an escape   fraction of 5\%.
This  value is  consistent  with  the  upper limits   derived from  the  Hopkins
Ultraviolet Telescope  (HUT)   by  Leitherer   et  al  (1995)   and  Hurwitz  et
al.  (1997)  and  more  recently   from   the   Far  Ultraviolet   Spectroscopic
Explorer   (FUSE) by Deharveng et  al. (2001).

At the intermediate redshifts $z\sim  1$ recent preliminary results by  Ferguson
(2001) obtained by means  of deep HST imaging  with the Space Telescope  Imaging
Spectrograph  STIS  seem to  indicate  a slightly higher average upper limit for
the   UV   ratio   of   the  order   of  $f_{esc} \lesssim 18$\%.

All   these  attempts   made at   low-intermediate redshifts have failed so  far
to  derive    a   statistically     significant  detection   of  the   UV escape
fraction.   Steidel,  Pettini  \&  Adelberger (SPA) (2001) on    the  contrary
have recently  obtained    the    first  significant    detection  of      Lyman
limit flux  in     a  sample    of  29     $z\simeq  3.40\pm     0.09$     Lyman
break galaxies     (LBGs)  using    spectra     obtained     in     a      multi
object   spectroscopy    configuration   devoted        to    the       redshift
identification   of the      high    $z$    galaxy   sample.     The   detected
Lyman     limit    flux  corresponds      to        an    escape   fraction   of
$f_{esc}   =   65$\%.   A value  that   is    $3.6   -   13$  times higher  than
previous  upper limits. 

In the  same period  we have  also started  a  program  to measure the UV escape
fraction   by  means     of  deep   long    slit     spectra    of    individual
galaxies.  This approach  avoids  possible  biases  inherent  to  the  procedure
of  combining   spectra  of   several  different   galaxies. We  report here the
results from  two galaxies  observed at  slightly lower  redshift ($z\simeq    3
-3.3$) with  the  FORS2  low resolution  spectrograph  at   the ESO  Very  Large
Telescope (VLT) telescope  in  the spectral  range $3500-6500$ {\AA}.

\section{Observations}

Our galaxies are two among the bright  end of the Steidel  et al. (1996)  sample
of  $z\sim  3$  star  forming  galaxies  observable  from  Paranal.    They  are
identified   as     C2   and      D6   and   have   been    observed by  Pettini
et     al.    (1998)   with        an infrared  spectrograph  to         compare
the  Balmer/OII line    emission  with the      UV     continuum       emission.
They are  not   included  in    the   subsample used  by   SPA to   produce  the
composite   spectrum and   measure   the  UV  ionizing    fraction. The      two
galaxies      have  magnitudes  of   $R=23.55$   and   $R=22.88$  and  redshifts
$z=3.319$ and  $z=2.961$  for  C2 and    D6, respectively,  and   have different
properties     in  terms   of Lyman-$\alpha$     emission,    UV      shape  and
dust reddening  as  shown by Pettini et al. (1998).

The galaxies  were observed  in the  period July  - August  2000 with the  FORS2
instrument  at   the  Kueyen   VLT  telescope   in  service   mode.   The  total
exposure times  were  19429    sec  for  C2    and   16432  sec   for  D6.   The
targets were  observed in  general at  low zenithal  distances with  low airmass
(1.07 and 1.3  for C2 and D6) to minimize atmospheric absorption.

The  spectroscopic observations  were performed  in long  slit with   the  grism
300V+20. The  useful wavelength  coverage ranges  from 3500  {\AA} to about 6500
{\AA} and the   spectral resolution   is 420.   The extraction  and  calibration
of the spectra has   been performed   using the  context LONG  in the  ESO-MIDAS
package.

After the standard bias subtraction and flatfield normalization, the 2D  spectra
were  stacked  from   the  original  frames  and   an  optimal  extraction   was
performed.  The wavelength  calibration in  each frame  was used  to check   the
stability  of the  2D target   position before  the stacking.  We optimized  the
subtraction procedure of the sky/detector background which  is    of course  the
main     source    of    random   and    systematic  errors    for   the    flux
calibrated   extracted    spectra.   Because   of    background  variation    on
several scales on the chip, sky  subtraction  was  optimized differently in  the
UV and red  part of the spectrum.  In particular,  different regions  above  and
below  the  target   spectrum  were  selected    in  the   UV  and    red parts.
A   linear fit  was adopted   for    the   background   shape  perpendicular  to
the    dispersion direction. The   slope  of   the  linear  fit was    different
for  the UV   and red part of  the spectrum.

In   this  respect, it  should   be emphasized   the  advantage for  long   slit
spectra  to perform an   optimal   sky  subtraction. Indeed   weak  fluctuations
in   emission   or absorption   perpendicular  to the  dispersion  direction can
affect     the estimate     of  the    background  level.   For  this    reason,
wide    regions sufficiently       far      away     from  the      target were
selected     for  the      background     subtraction.    In  this  way,     the
weak  fluctuations present  in these  wide regions  were easily removed.

The  spectrum  extraction was  performed   using an  optimal   extraction method
fitting   the  spatial    profile of  the   spectrum. The extraction  slit   was
quite   short, 7      pixels  or   1.4 arcsec,        to     maximize        the
signal-to-noise ratio.

Standard  flux   calibration  was  performed  to    provide  a   relative   flux
calibration   of  the   spectra.  Wavelength  calibration was performed applying
the  dispersion  relation  without rebinning   at any  constant  wavelength  bin
size.   This  procedure  keeps  unchanged    the  original    pixel  noise   but
produces  a slight variation of the wavelength bin along the spectra.

The calibrated spectra were  corrected for atmospheric  extinction  adopting the
airmass values previously mentioned.  Finally they have been corrected  also for
interstellar extinction using  the E(B-V) values   of  0.015 and   0.05 obtained
from the NASA/IPAC Extragalactic Database  for    the  galaxies  D6    and   C2,
respectively  and    adopting  the standard Seaton (1979) extinction curve.

The  final extracted  and  calibrated  spectra are  shown  in  Fig.1 and  Fig.2
(upper panels)  with  the  resulting noise     level. Strong  SiII1260,  OI1303,
CII1334, SiIV1398,  SiII1526,  CIV1548 absorption   lines   are clearly  visible
in  the spectra  and   give  the   same  absorption   redshifts   for  the   two
galaxies  as provided  by  Pettini  et  al. (1998). Their rest  frame equivalent
widths  are  of  the order  of  3   {\AA}. At   the  same    time,  the   
Lyman-$\alpha$    emissions   are  different     being   stronger    in   D6      with
$W=41/(1+z)=10.5$   {\AA}   than   in   C2,    $W=14.5/(1+z)=3.4$   {\AA}    and
correspondingly   the  UV  dust  absorption estimated by  Pettini et  al. (1998)
in  the  two   galaxies  is   stronger in   C2 (A$_{1500}$=1.14)   than in   D6
(A$_{1500}$=0.8).

\section{Results}

It is clear  from Fig.1 and  Fig.2 (lower panels)  that there is  no significant
emission shortward of the Lyman limit (912 {\AA}) in both galaxies. The  average
flux is zero and the noise level in the spectra can provide a useful upper limit
to the  $f_{1500}/f_{900}$ ratio  and hence   to the  UV escape  fraction. As in
SPA  we define a spectral  region shortward of the   Lyman Limit
sufficiently  large to have  enough statistics  but keeping at the same  time  a
low noise  level. The selected rest  frame region is between 880 and 910 {\AA}.

The  pixel-to-pixel noise  rms obtained  in the  selected region  is     $\sigma
(880-910)=3.3$  (relative  flux  densities  in   Hz$^{-1}$)  for   the    galaxy
C2 which  corresponds     to    an    average      upper    limit   flux      in
the region  of   $f_{900}<3.3/\sqrt{54  pxl}=0.45$.  The  corresponding  average
flux level  at 1500   {\AA}     is   $33\pm      2.5/\sqrt{21  pxl}       =33\pm
0.6$.    The    resulting     lower      limit      to      the         observed
$f_{1500}/f_{900}$ ratio is $f_{1500}/f_{900}>73$ for the galaxy C2. 

Similarly for the  galaxy D6  the   pixel-to-pixel noise  rms  obtained   in the
selected  region  is     $\sigma (880-910)=1.1$  corresponding to  an    average
upper     limit    flux      in   the       region   of    $f_{900}<1.1/\sqrt{50
pxl}=0.16$.  The  corresponding   average   flux level  nearby 1500  {\AA}  rest
frame is  $11\pm 1.5/\sqrt{30 pxl}       =11\pm   0.3$.   The   resulting  lower
limit to the observed $f_{1500}/f_{900}$ ratio is $f_{1500}/f_{900}>69$.

The escape fraction defined  as the fraction of  emitted 900 {\AA} photons  that
escapes the  galaxy without  being absorbed  is difficult  to estimate  since it
depends on  the intrinsic  spectral shape  of the  source and  its reddening.  A
definition of the UV ionizing  escape fraction $f_{esc}$  related to  the actual
measure  is   represented  by  the  ratio   $f_{1500}/f_{900}$  related  to  the
1500 {\AA} escape fraction    by   SPA    to   avoid     any  dependence      on
uncertain  parameters  like       dust    reddening.           Specifically   we
adopt   a relative    escape       fraction     $f_{rel,esc}     \equiv        3
\times f_{900}/f_{1500}$ where       the    factor   3  is        the  $f_{1500}
/  f_{900}$     ratio   expected         from        the    stellar    synthesis
models (cf.    SPA).   The uncertainties  on    this   predicted  value   depend
on   the adopted  initial mass function and stellar ages and ranges from   about
2.5   to 5.5.  We  adopt    the same  factor  3   used by   SPA  for comparison.

To use the observed $f_{1500}/f_{900}$  ratios for the estimate of  the relative
UV  escape fraction, a  correction  should be  made  to  take into  account  for
the  opacity present  shortward of the  Lyman-$\alpha$ and in  particular in the
Lyman continuum region.  This opacity  is due   to   the   intervening   neutral
hydrogen   not directly   associated  with  the   galaxies,   the   so    called
intergalactic  medium   (IGM).     This  correction    is  negligible    at  low
redshifts  $z<0.1$ but    becomes   important  at   $z\sim     3$  and    should
be   taken    into account.    It  is     clear that    a careful     correction
for       intervening  absorption    in   individual   objects  requires    high
resolution  spectra.     In our    case  we     adopt   the     same  correction
estimated   by    SPA in   an empirically  way using a  composite   QSO spectrum
with   a  mean    redshift     of  $z\simeq   3.47$.  From     this    composite
spectrum   the  decrement    due    to intervening   HI absorption     has  been
estimated     to be  $f_{1500}/f_{900}  =  3.85$.   Since   our  galaxies   have
redshifts   not   far  from  that  of the    QSO composite   spectrum, we   have
adopted the same  correction factor.

The  resulting   $1\sigma$  upper  limit  to  the  relative  UV  ionizing escape
fraction in    both galaxies      at   $z=3-3.3$    is  $f_{rel,esc}\equiv     3
\times    3.85\times f_{900}/f_{1500}<    16$\%.

\section{Discussion}

Steidel and  collaborators have provided  the first  significant detection in  a
sample   of   $z\simeq  3.4$  Lyman   break  galaxies  (LBGs). They have used  a
composite   spetrum   made   by  29   LBGs   spectra   obtained   from   the Low
Resolution  Imaging   Spectrometer  at the  Keck  telescope.   For  the  average
sample they derive  an  observed ratio   $f_{1500}/f_{900} =   17.7\pm  3.8$. In
Fig.1  and    Fig.2  (lower   panels)  the   expected    ratio  derived  by  SPA
corresponds  to    the  thick  horizontal   line   shown  between  880 and   910
{\AA}. The smoothed   versions of the   spectra to the   instrumental resolution
are  also   shown    as  thick    curves    only   for   illustrative  purposes.
The  smoothed spectra fluctuate  around the zero flux level  shortward of    910
{\AA}    and    the     resulting  upper    limit    on    the relative   escape
fraction,  $f_{rel,esc}\lesssim  16$\%,   is 4      times     lower      than    the
$f_{rel,esc}=65$\% SPA detection.

This result  can also be  compared with other  measures derived  for  individual
galaxies  at  lower  redshifts.  At  $z<0.1$  the  most  recent  result has been
obtained  by  Deharveng et   al. (2001)   for the  nearby  galaxy  Mrk 54  using
the {\it Far Ultraviolet Spectroscopic   Explorer (FUSE)}. No flux was  detected
shortward  of  910  {\AA}  and  the lower  limit  to the  ratio  of $f_{1500}  /
f_{900}$,  as  measured  from  the  noise  level,  was  108  (converting   their
fluxes  in     erg/s/cm$^2$/Hz)  without   correction  for     intervening   IGM
absorption). This  can  be  translated  in  an  upper limit   to  the   relative
UV  escape   fraction  of   $f_{rel,esc}  \lesssim     3$\%  (this   value    is
different from the one reported    by   the   authors     because   of     their
different definition   of the       escape  fraction      which   depends     on
the     flux emitted     in  the H$\alpha$   line).  Other   4     $f_{1500}   /
f_{900}$ ratios   reported in  their   paper     from  previous     observations
with the       {\it  Hopkins  Ultraviolet   Telescope}  (Leitherer     et    al.
1995) give      average  upper   limits    in      the    range     $f_{rel,esc}
\lesssim 20-30$\%     (with no correction   for  intervening  IGM absorption).

At the intermediate redshifts $z\sim  1$ recent preliminary results by  Ferguson
(2001) obtained by  means  of deep  HST imaging  with  the {\it Space  Telescope
Imaging Spectrograph   (STIS)}  seem  to   indicate   a slightly  higher average
upper limit  for  the    UV    ratio    of    the   order    of   $f_{1500}    /
f_{900}  \sim  17$    corresponding  to    a   relative   escape   fraction   of
$f_{rel,esc}  \lesssim  18$\% (preliminary  estimate  including intervening  IGM
absorption).

Our upper limits  derived from the  individual spectra of  two galaxies at  $z=3
-3.3$ are more  similar to the  limits derived from  HUT, FUSE and  HST at lower
redshifts  than to  the detection  by  SPA  using a  composite Keck
spectrum of galaxies at $z\sim 3.4$.

In an attempt  to understand the  difference between the  result by SPA  and our
result, we note the following. The upper limit for the two galaxies we  observed
are  very   close   in   value,     although    the    two     galaxies     have
different  properties      in   terms   of      Lyman-$\alpha$     emissions and
dust reddening.   On    the     other  hand     it     is    clear      that the
Lyman-$\alpha$  emissions   of      our  galaxies  ($W=    3, 11$     {\AA}  for
C2,   D6  respectively)   are    lower      than  the     average Lyman-$\alpha$
emission in   the SPA  composite   spectrum  ($W\sim   20$   {\AA}).  In   fact,
the $z\simeq   3$  galaxy     sample used    by    Shapley      et   al.  (2001)
to  estimate    the        optical luminosity    function  shows       a  median
Lyman-$\alpha$   equivalent   width of    0.  This    implies  that about   half
of  the  sample     shows   Lyman-$\alpha$   in   absorption  and  half    shows
Lyman-$\alpha$    in    emission. The  Lyman-$\alpha$    equivalent  widths   of
our  two       galaxies   are    close   to    the  typical    $W$     of    the
spectroscopic galaxy sample   at  $z\sim    3$. Thus,   it  is   possible   that
significant  UV ionizing    photons  can   escape   only  the  fraction of   the
$z\sim  3$ galaxy population with  strong Lyman-$\alpha$ emission. To test  this
hypothesis it could be interesting to compare the escape fraction for the strong
Lyman-$\alpha$  emitters  to  that  for  the  galaxies  with  Lyman-$\alpha$  in
absorption in the SPA sample.

If this hypothesis were correct it would imply that the average escape  fraction
of ionizing UV   photons present  at   $z\sim  3-3.5$  could   be  smaller  than
previously  thought.  The consequences  of  this  for the  estimate of   the UVB
produced by  the  high  $z$ star   forming galaxies  can  be  evaluated adopting
as  a  reference  value  our upper  limit    of  $f_{rel,esc}   \lesssim  16$\%.  At
$z\sim  3$   the    galaxy contribution    has   been   recently  evaluated   by
SPA   who   adopted their  estimate   of  the   1500  {\AA} galaxy    luminosity
function    and their derived escape   fraction of  65\%.  Their   resulting UVB
intensity from the   galaxy population   alone    is   $J=1.2\times    10^{-21}$
ergs   s$^{-1}$  cm$^{-2}$ Hz$^{-1}$   sr$^{-1}$  which    is  already    larger
than    the   value   estimated  by     the   "proximity     effect"   in    the
Lyman-$\alpha$  forest   of    QSO absorption   spectra,  $J=5-7\times 10^{-22}$
ergs   s$^{-1}$ cm$^{-2}$ Hz$^{-1}$ sr$^{-1}$ (see   e.g. Giallongo  et      al.
1996; Scott    et  al.  2000).  The addition  of    the   contribution   by  the
QSO  population    would  increase  this   discrepancy.  To    alleviate     the
problem  they   argued   about    a possible   overestimate  of        the   QSO
contribution   to        the   UVB. Bianchi          et     al.     (2001)  used
the        observed    UV            star   formation          history       and
three    different           values for     the  intrinsic $f_{1500}   /f_{900}$
ratio,     namely        5.3,    21,  42  corresponding  to         $f_{rel,esc}   =
57,      14,     7$\%,      respectively,  (following     the    definition   of
$f_{rel,esc}$  adopted        in     the       present  paper)   to   compare    the
contribution  of  the two    populations  to    the    UVB  and the  effects  on
the  evolution  of  the Lyman-$\alpha$    forest.     It    is found       that,
to      avoid        an   overprediction      of    the   total (QSO$+$galaxies)
UVB    respect     to   the  value   derived    from    the "proximity  effect",
the relative escape fraction  should be of  the order of or smaller than   15\%.
In   this respect   our upper   limit supports  a scenario where the   total UVB
is consistent  with    the "proximity effect"   at  $z\sim 3$   with a    galaxy
contribution    of   $J\lesssim     6\times  10^{-22}$  ergs s$^{-1}$  cm$^{-2}$
Hz$^{-1}$ sr$^{-1}$,  i.e.  no    more   than  a   factor     2.5 higher    than
the   QSO  contribution.   An   UVB    produced    by    this  mixed  population
could    account  for    the  metal  enrichment   of    the  IGM    and for  the
SiIV/CIV    metal-line     ratios   observed   in    high       resolution   QSO
absorption  spectra  (see  e.g.   Giroux \& Shull 1997).

In summary,  many pieces  of evidence,  from the  limits derived  at lower   and
intermediate redshifts to  the upper limit of $f_{rel,esc} \lesssim 16$\%  presented
in  the   present paper,  point toward   a lower   ionizing escape  fraction. We
emphasize that our   result, however robust,   is based on   the observations of
two $z\sim 3$ galaxies  only. A large  statistical sample is  needed to  confirm
its implication on the UV background evolution. This is within the grasp of  the
current instrumentation at very large telescopes.

\acknowledgments

We are grateful to C. C. Steidel, M. Pettini and collaborators for providing the
finding charts of  the two galaxies. 

\bigskip


\noindent

\newpage

\begin{figure}
\plotone{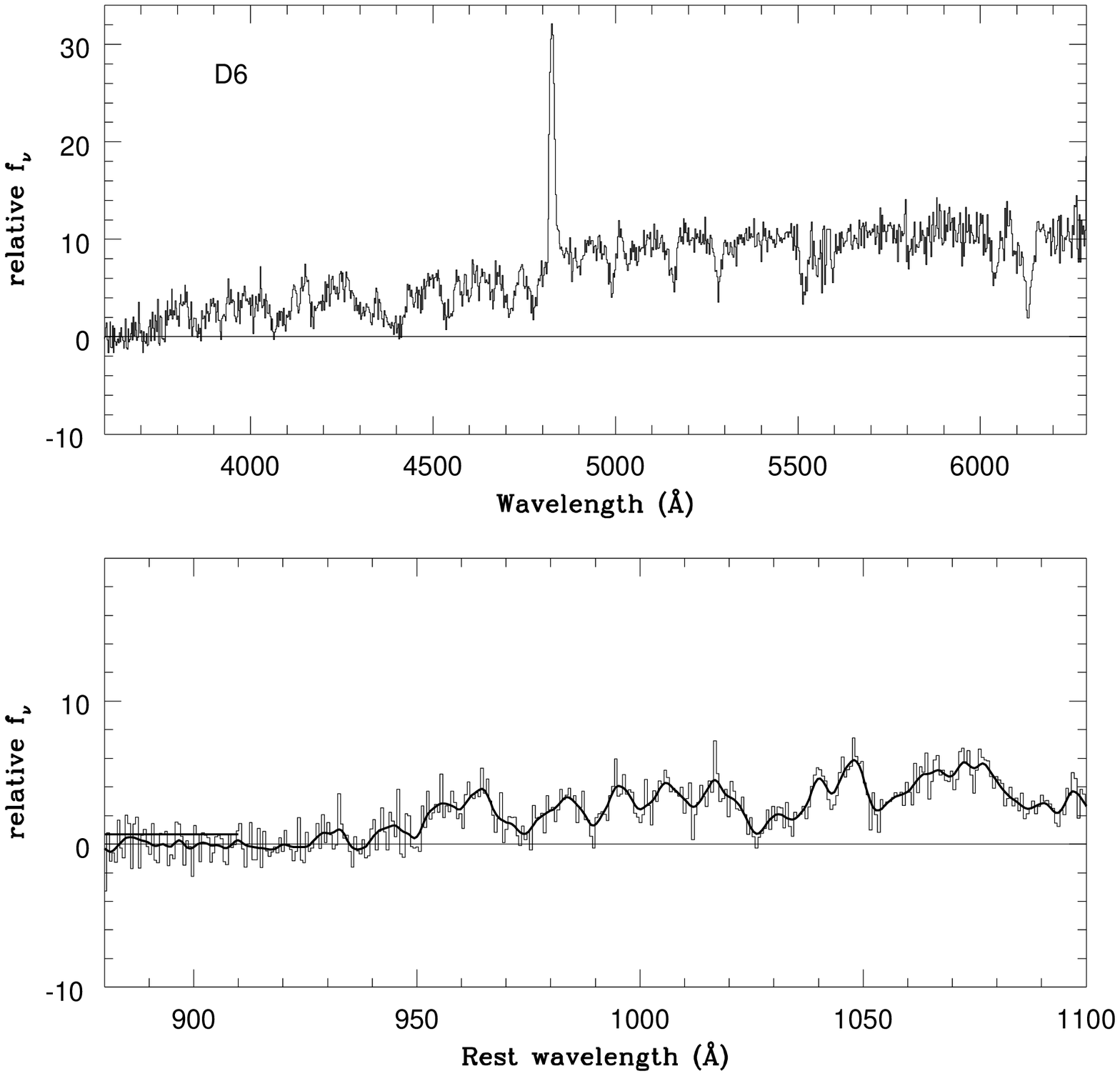}
\caption{
The spectrum of the galaxy D6 is shown as a function of
the observed  (upper panel)  and the  rest frame  (lower panel) wavelengths. The
thick curve  in the  lower panel  is the  smoothed version  to the  instrumental
resolution. The thick horizontal  line from 880 and  910 {\AA} is the  predicted
Steidel et al. level in the spectrum.\label{fig1}
}
\end{figure}

\newpage

\begin{figure}
\plotone{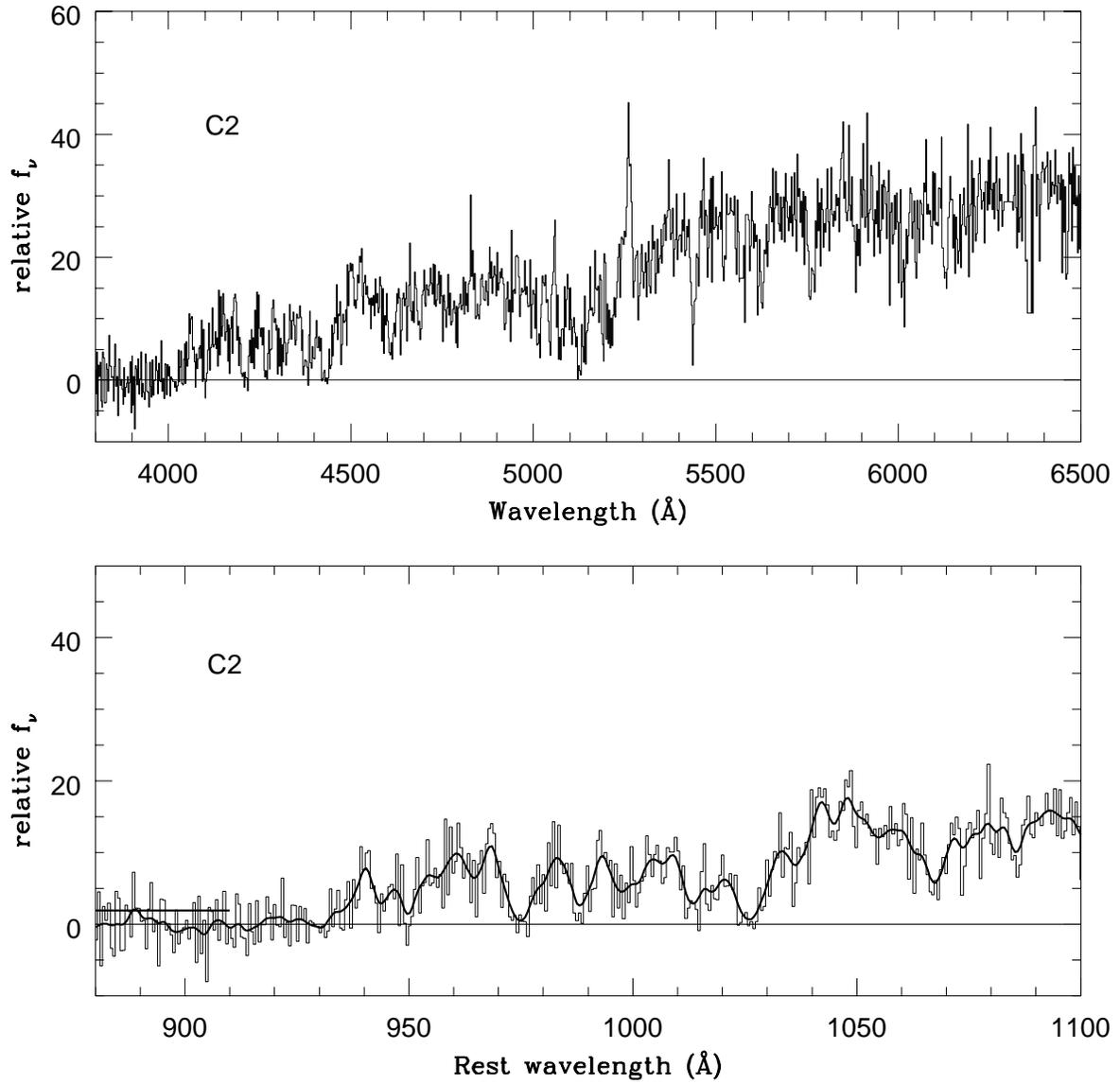}
\caption{
The spectrum of the galaxy C2 is shown as in Fig.1.
\label{fig2}
}
\end{figure}

\end{document}